\documentclass[aps,prb,twocolumn, showpacs, amsmath, amssymb, 
superscriptaddress]{revtex4-1}
\usepackage{graphicx}
\usepackage{dcolumn}
\usepackage{bm}
\usepackage{natbib}
\newcommand{\BCPO}{\ensuremath{\text{BiCu}_{2}\text{PO}_{6}\,\,}}

\newcommand{\degree}{\ensuremath{^{\circ}}\,\,}
\newcommand{\figref}[1]{Fig.\,\,\ref{#1}}
\newcommand{\CuIon}{\ensuremath{\text{Cu}^{\text{2+}}\,}}
\newcommand{\rvec}[1]{\ensuremath{\textbf{#1}^{*}}}

\begin{document}
\title{ Incommensurate dynamic correlations in the quasi-two-dimensional spin
liquid \BCPO }
\author{K.W. Plumb}
\email{kplumb@physics.utoronto.ca}
\affiliation{Department of Physics, University of Toronto, Toronto, Ontario M5S
1A7, Canada}

\author{Zahra Yamani}
\affiliation{Canadian Neutron Beam Centre, National
Research Council, Chalk River Laboratories, Chalk River, Ontario, K0J
1P0, Canada}

\author{M. Matsuda}
\affiliation{Quantum Condensed Matter Division, Oak Ridge National Laboratory,
Oak Ridge Tennessee 37831, USA}

\author{G. J. Shu}
\affiliation{Center for Condensed Matter Sciences, National Taiwan University,
Taipei, 10617 Taiwan}

\author{B. Koteswararao}
\affiliation{Center for Condensed Matter Sciences, National Taiwan University,
Taipei, 10617 Taiwan}

\author{F.C. Chou}
\affiliation{Center for Condensed Matter Sciences, National Taiwan University,
Taipei, 10617 Taiwan}

\author{Young-June Kim}
\email{yjkim@physics.utoronto.ca}
\affiliation{Department of Physics, University of Toronto, Toronto, Ontario M5S
1A7, Canada}

\date{\today}
\begin{abstract}
We report detailed inelastic neutron scattering measurements on single crystals
of the frustrated two-leg ladder \BCPO\!\!, whose ground state is described as 
a spin liquid phase with no long-range order down to 6~K. Two branches of
steeply dispersing long-lived spin excitations are observed with excitation
gaps of $\Delta_1 = 1.90(9)$~meV and $\Delta_2 = 3.95(8)$~meV. Significant
frustrating next-nearest neighbor interactions along the ladder leg drive the
minimum of each excitation branch to incommensurate wavevectors  $\zeta_1 =
0.574\pi$ and $\zeta_2 = 0.553\pi$ for the lower and upper energy branches 
respectively. The temperature dependence of the excitation spectrum near the 
gap energy is consistent with thermal activation into singly and doubly 
degenerate excited states. The observed magnetic excitation spectrum as well as
earlier thermodynamic data could be consistently explained by the presence of
strong anisotropic interactions in the ground state Hamiltonian.
\end{abstract}

\maketitle

\section{Introduction}
Low-dimensional quantum antiferromagnets can realize a rich and diverse array
of physical phenomena from a seemingly simple set of interactions.
The canonical low-dimensional quantum antiferromagnet is the spin-1/2
Heisenberg chain, in which neighboring spins are coupled with antiferromagnetic
exchange $J_1$. The system does not have any long range order and the spin-spin
correlations decay with a power law. In spin-1/2 chains the elementary
excitations are S = 1/2 quasiparticles termed spinons and the dynamic
susceptibility is dominated by a gapless dispersive continuum.
\cite{Tennant:95,Dender:96} Frustration can be introduced to the spin-1/2 chain
by competing antiferromagnetic next-nearest-neighbor (NNN) interaction $J_2$.
For a large NNN exchange of  $J_2/J_1 > 0.241$ the ground state is composed of
dimerized singlets \cite{Haldane:82,Okamoto:92} and the excitation spectrum is
described with a coherent triplet band of gapped excitations. At the
Majumdar-Ghosh (MG) point $ J_2/J_1 = 0.5$ the singlet ground state is exact
and the system forms a one-dimensional dimer crystal.\cite{Majumdar:69} As the 
NNN exchange is increased beyond the MG point frustration drives the
spin correlations in the dimerized system to an incommensurate wavevector.
\cite{Bursill:95,White:96}

A gap may also be introduced into the excitation spectrum by coupling two
neighboring chains via a short range interaction $J_{rung}$, to create an even
leg ladder \cite{Dagotto:96}. The rung coupling confines spinons into S=1
magnons and the low energy dynamic susceptibility exhibits a well defined,
triply degenerate, single particle peak.\cite{Shelton:96} Theoretical
understanding of the properties of even-leg ladders is highly developed
\cite{Gopalan:94,Shelton:96,Giamarchi:99} and a number of experimental
realizations have allowed for a detailed understanding of the very rich
physical phenomena present in spin-ladders for both the strong rung coupling
(rung-singlet) \cite{Masuda:06,Garlea:07,Thielemann:09} and strong-leg coupling
(Haldane) \cite{Hong:10,Schmidiger:12} regimes. However, the physics of spin
ladders with additional frustrating NNN interactions is not understood well.
Field theoretical and numerical investigations have indicated that the addition
of frustration can lead to many exotic quantum ground states not realized in
the standard two-leg ladder,\cite{Nersesyan:98,Vekua:06,Lavarelo:11} yet
comprehensive experimental investigations are still lacking.

\begin{figure}[htb]
    \includegraphics[]{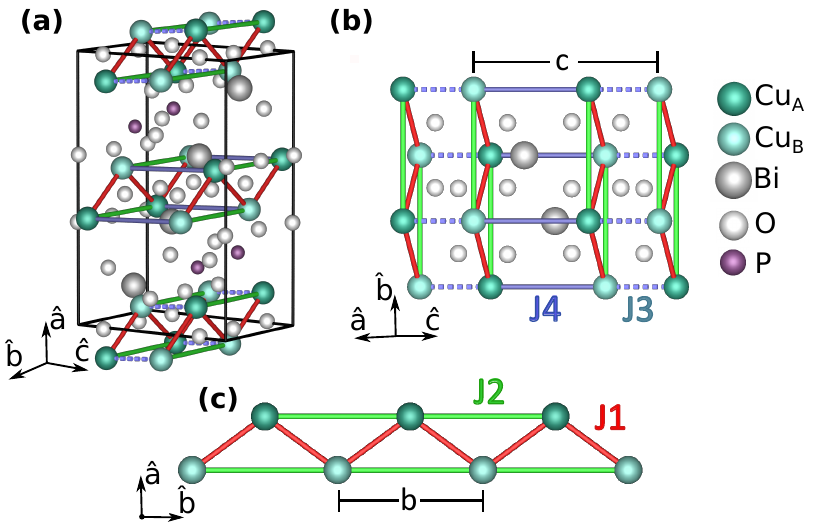}
    \caption{\label{fig:structure} (a) Schematic representation of the crystal
    structure of \BCPO\!\!. The unit cell is orthorhombic, space group Pnma with
    a = 11.755 \AA, b = 5.16 \AA, c = 7.79 \AA\,\,at 6~K.\cite{Mentre:06} (b) 
    Perspective view of the ladder unit in the b-c plane and (c) projection 
into the a-c plane showing the zigzag chains of \CuIon ions.}
\end{figure}

Such a frustrated two-leg ladder seems to be realized in a newly discovered
quasi-two-dimensional spin-1/2 compound 
\BCPO\!\!.\cite{Koteswararao:07,Mentre:09} The crystal structure of
\BCPO is shown in \figref{fig:structure}.  Zig-zag chains of \CuIon ions run
parallel to the crystallographic b-axis, as shown clearly by a projection in the
$\mathbf{a}-\mathbf{b}$ plane in \figref{fig:structure} (c).  Along the chains
\CuIon ions at crystallographically inequivalent sites, labeled
$\textrm{Cu}_\textrm{A}$ and $\textrm{Cu}_\textrm{B}$, interact via NN
antiferromagnetic exchange $J_1$ and NNN antiferromagnetic exchange $J_2$.  The
chains are coupled along the c-axis by both $J_3$ and $J_4$.
Although the $J_3$ interaction distance is shorter than $J_4$, band structure
calculations predict $J_4$ to be the dominant coupling along the
c-axis.\cite{Tsirlin:10} Structurally, this is a result of the particular
superexchange pathways. $J_3$ is a nearly 90\degree Cu-O-Cu bond, which is
usually small and ferromagnetic, in contrast $J_4$ is mediated by a nearly
180\degree  Cu-O-O-Cu bond.  Therefore, \BCPO can be considered as a system of
$J_1$-$J_2$-$J_4$ ladders with a weaker interladder coupling $J_3$.
\cite{Mentre:09,Tsirlin:10}

Magnetic susceptibility, magnetization, and heat capacity measurements on
powder samples have consistently reported that \BCPO has a singlet ground state
with a spin excitation gap of $\sim$2.9 meV.
\cite{Mentre:06,Koteswararao:07,Koteswararao:09,Tsirlin:10} The thermodynamic 
measurements cannot be interpreted in terms of simple gapped one-dimensional 
(1D) models, including the $J_1$-$J_2$ chain and the two-leg ladder, but 
require a combination of frustration and two-leg-ladder 
geometry.\cite{Koteswararao:09,Mentre:09} Inelastic neutron scattering (INS) 
measurements on powder samples have reported that \BCPO has a dispersive spin 
excitation spectrum with a gap of $\Delta = 2.9$~meV and that significant 
frustration is required to describe the powder averaged structure 
factor.\cite{Mentre:09} More recent high field magnetization measurements on 
single crystals have revealed that the critical field for the closing of the 
spin gap is dependent on the direction of the applied field indicative of 
strong anisotropic interactions in the magnetic Hamiltonian.\cite{Kohama:12} 

In real materials (usually) small anisotropic interactions, additional to the
Heisenberg exchange couplings, invariably exist.  Such interactions are often a
negligible perturbation to the isotropic Heisenberg interaction, as in cuprate
superconductors\cite{Keimer:92:1,Keimer:92:2}, but can sometimes alter the 
magnetic properties in a
fundamental way. As an example, for compounds that lack local inversion
symmetries anisotropic Dzyaloshinksky-Moriya 
(DM)\cite{Dzyaloshinsky:58,Moriya:60} interactions are permitted in
the magnetic Hamiltonian. Depending on the particular crystallographic
symmetries the DM interaction can induce a staggered field.\cite{Affleck:99} In
1D chains this staggered field results in an effective confinement potential
between spinons and the appearance of incommensurate gapped modes on application
of a magnetic field.\cite{Affleck:99} The staggered field also has a drastic
effect for spin ladders. In contrast to a spin ladder in a uniform magnetic
field, which transitions to a gapless phase above a critical field, the presence
of a staggered field is predicted to lead to a quantum phase transition between
two gapped phases above and below the critical field.\cite{Wang:02}

The numerous exchange pathways in \BCPO complicate interpretation of 
thermodynamic data and important details including interladder coupling and 
anisotropic interactions are often neglected in the analysis. In order to 
determine the microscopic spin Hamiltonian of a complex magnetic system with 
competing interactions and anisotropic exchange couplings, INS measurements 
using a single crystal sample are essential.

We have conducted an extensive neutron scattering study of the magnetic 
excitation spectrum in \BCPO\!\!. Our data confirms that \BCPO is appropriately 
described by weakly interacting two-leg ladders with incommensurate dynamic 
correlations driven by frustration.  In contrast to the single triply 
degenerate excitation branch expected for an isotropic ladder, we observe two 
branches of steeply dispersing, long-lived, excitations.  The excitation gap of 
each mode, directly probed by INS, was measured to be $\Delta_1 = 1.90(9)$~meV 
and $\Delta_2 = 3.95(8)$~meV, this differs significantly from the $\sim2.9$~meV 
gap extracted from thermodynamic measurements 
\cite{Mentre:06,Koteswararao:07,Tsirlin:10} indicating that current models are 
inadequate to describe the ground state of \BCPO\!\!.  Furthermore, the 
temperature dependence of the incommensurate modes are consistent with a 
description in terms of thermal activation into singly and
doubly degenerate modes. We argue that, in addition to frustration, strong
anisotropic interactions are important for understanding the physics of this
material.

\section{Experimental Details}
Experiments were carried out on a 4.5~g single crystal sample grown using the
traveling floating zone method. The sample mosaic was measured by neutron
scattering to be $0.2\degree$\!at T = 6~K. Measurements were performed on the
C5 DUALSPEC triple axis spectrometer at the Canadian Neutron Beam Centre at 
Chalk River Laboratories and on the HB1 triple axis spectrometer at HFIR. Both
instruments employed a vertically focusing pyrolytic graphite (PG) 
monochromator and C5 was equipped with a flat graphite analyzer while HB1 
utilized a fixed vertically focusing analyzer. On C5 the sample was mounted in 
the $(0,k,l)$ scattering plane and the spectrometer was operated at a fixed 
final energy of 14.56~meV. Experiments on HB1 were performed with the sample 
mounted in the $(h,k,2k)$ horizontal scattering plane at a fixed final energy 
of 14.7~meV. Temperature control was provided by a closed cycle cryostat. All 
data was corrected for higher-order wavelength neutrons in the incident beam 
monitor.\cite{Stock:04} Intensities were placed on an absolute scale by 
normalization with the integrated intensity of a transverse acoustic phonon 
measured near the $(0 0 4)$ Bragg peak on the respective instrument.

The crystallographic unit cell of \BCPO is orthorhombic, space group Pnma with 
a = 11.755~\AA, b = 5.16~\AA, c = 7.79~\AA. Along the zig-zag chains NN \CuIon 
ions in \BCPO are separated by b/2 so that the momentum of magnetic excitations 
in the $\mathbf{b}^{*}$ direction is indexed using $\mathbf{\tilde{k}} = 
\mathbf{q}\cdot \mathbf{b}/2 =  \pi \mathbf{k}$.

\section{Experimental Results}
\subsection{Spin excitation spectra}
\begin{figure}[htb]
    \includegraphics[scale = 0.9]{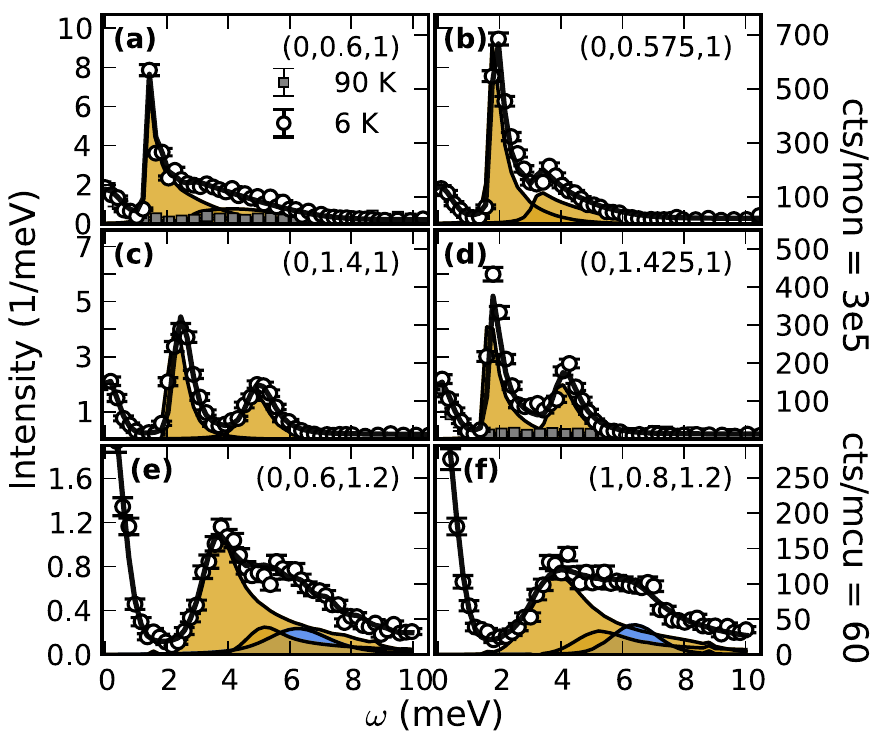} 
    \caption{\label{fig:const_Q_scans} Representative constant-\textbf{Q} 
    scans.  Data in panels (a)-(d) were obtained on C5 with the sample aligned 
    in the $(0,k,l)$ scattering plane and a collimation of 33'-48'-51'-144'.  
    Data in panels (e) and (f) was obtained on HB1 with the sample aligned in 
    the $(h,k,2k$) scattering plan and a collimation of 48'-40'-40'-120'.  All 
    data was collected at T = 6~K. The solid black lines are the results of a 
global fit to the single mode approximation, gold and blue filled areas show 
the contribution from each mode including resolution effects.}
\end{figure}

\begin{figure*}[htb]
    \includegraphics[scale = 0.9]{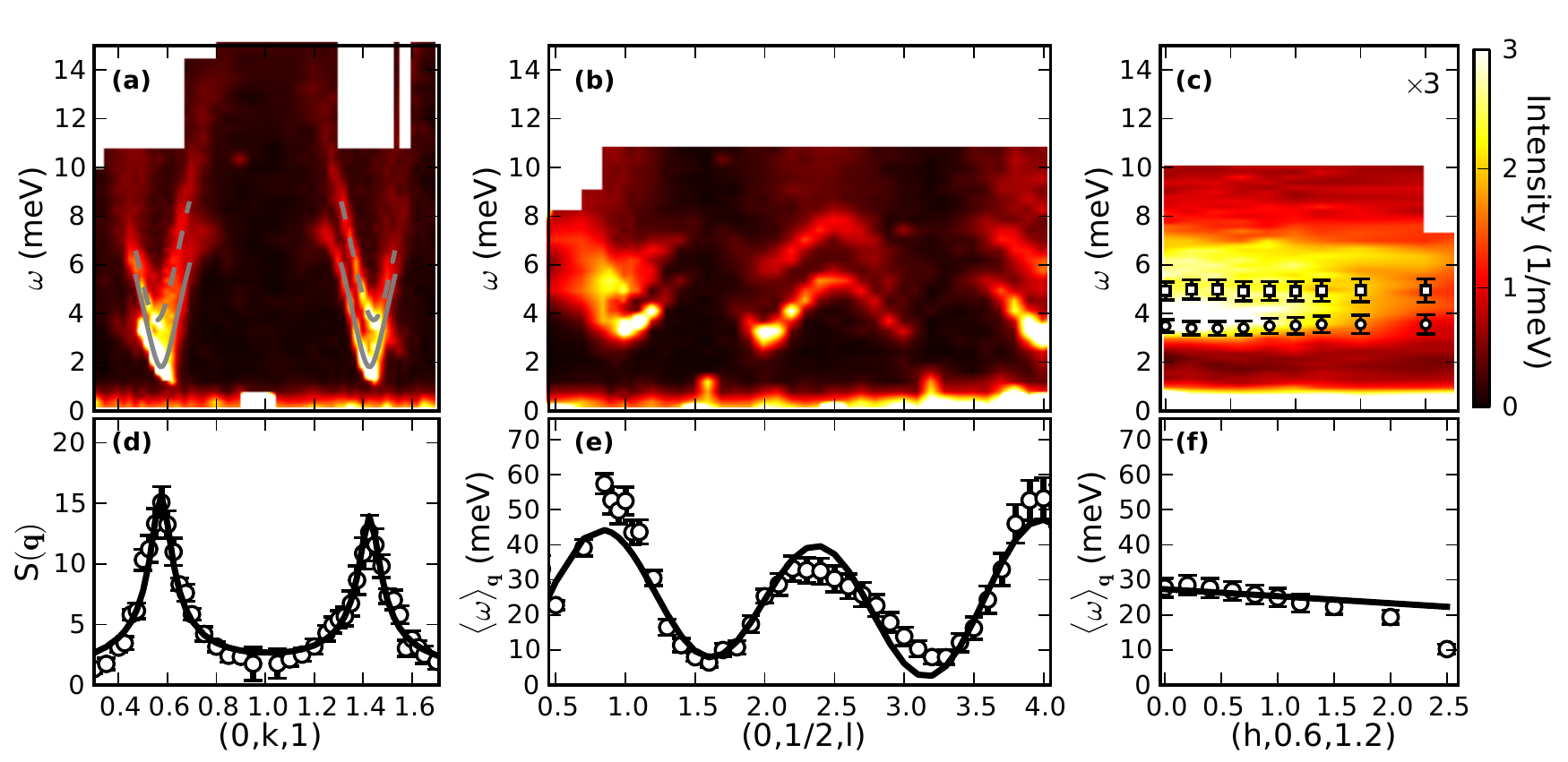}
    \caption{\label{fig:SQW_maps}(a) - (c) Inelastic neutron scattering
    intensity map for T = 6K constructed by linear interpolation of a series of
    constant scans. All data have been corrected for the isotropic \CuIon form
    factor.  Solid and dashed gray lines show the dispersion relation, Eq.\!
    \eqref{eq:disp}, using the parameters obtained from the single mode
    approximation fit. Solid white symbols in (c) show the fitted peak 
    positions of resolution limited mode. (d) Structure factor data obtained at 
    T = 6 K, for equal-time spin correlations along the chain \rvec{b} 
    direction, calculated by numerically integrating constant \textbf{q} scans 
    for $\omega > 0.5$ meV. (e) - (f) Measured first moment as a function of 
    wavevector along the rung \rvec{c} direction and  perpendicular to the 
    planes of the ladders, along \rvec{a}. Solid lines are fits as described in 
the text.}
\end{figure*}

The momentum and energy dependence of spin excitations in \BCPO were surveyed 
through a series of constant-\textbf{Q} scans. No evidence for elastic magnetic 
scattering at T = 6~K was found indicating the absence of static magnetic order 
in \BCPO\!\!. Representative scans taken in proximity of the spin gap at T=6~K 
are shown in \figref{fig:const_Q_scans}.
Throughout most of the Brillouin zone the scattering intensity is dominated by 
two well defined and highly dispersive modes. The dispersion of these modes is 
not commensurate with the structural unit cell and the scattering intensity 
vanishes as the temperature is increased above 60~K, where a broad maxima in 
the magnetic susceptibility was observed\cite{Koteswararao:07}, confirming the 
magnetic origin.  Extensive sampling of \textbf{Q}-values throughout reciprocal 
space enabled the construction of a map of the dynamic structure factor 
$S(\mathbf{q},\omega)$ shown in \figref{fig:SQW_maps} (a) - (c).

Magnetic excitations in \BCPO are highly anisotropic, with a bandwidth of
$\sim$12 meV in the $\mathbf{b}^{*}$ direction and $\sim$2 meV in the
$\mathbf{c}^{*}$ direction. Dispersion along the $\mathbf{c}^{*}$ direction 
confirms the presence of a sizable interladder interactions. Any dispersion 
along the $\mathbf{a}^{*}$ direction could not be resolved by our thermal 
triple axis measurements, indicating that \BCPO should be regarded as a
quasi-two-dimensional system.

Within the first Brillouin zone, each branch of the excitation spectrum has two 
minima occurring at wavevectors $\zeta_{\alpha} = 0.5 + \delta_{\alpha}$ and 
$\zeta_{\alpha} = 1.5 - \delta_{\alpha}$. The distinct double-well structure is 
reminiscent of the dispersion in $J_1$-$J_2$ chains and the incommensurate 
minima is indicative of a strong competition between $J_1$ and 
$J_2$.\cite{Bursill:95,White:96,Nersesyan:98,Lavarelo:11}

The excitation spectrum shown in \figref{fig:SQW_maps}(a) cannot be described
using strong-coupling expansions for a two-leg ladder with competing
interactions.  \cite{Tsirlin:10,Lavarelo:11} These perturbative expansions
require that the rung coupling ($J_4$) is much larger than the coupling along
the chains ($J_1$,$J_2$).  The large bandwidth of each branch compared with the
gap energies implies that \BCPO likely lies in an intermediate coupling regime
where $J_1 \sim J_4$, this is in qualitative agreement with band-structure
calculations.\cite{Tsirlin:10}

Lacking a microscopically motivated theory for \BCPO\!\!, the essential features
of the lowest energy excitations were determined by fitting data in the vicinity
of the dispersion minimum with a single mode approximation (SMA) cross-section
and an empirical dispersion relation:
\begin{align}
    \left(\omega^{\alpha}_{\mathbf{q}}\right)^2 = \Delta_{\alpha}^2 +
    v_{\alpha}^2\sin^2\left(\tilde{k} - \pi \zeta_{\alpha}\right),
   \label{eq:disp}\\
   S(\mathbf{q},\omega) = \frac{1}{1 - e^{-\omega^{\alpha}_{\mathbf{q}}/T}}
   \frac{\mathcal{A}^{\alpha}_{\mathbf{q}}}{\omega^{\alpha}_{\mathbf{q}}}
   \delta(\omega - \omega^{\alpha}_{\mathbf{q}}),
    \label{eq:SMA}
\end{align}
$\mathcal{A}^{\alpha}_{\mathbf{q}}$ is a mode dependent intensity pre-factor, 
$v_{\alpha}$ parameterizes the spin-wave velocity of each mode, and we set 
$\hbar = k_B = 1$ throughout this paper.  Equation \eqref{eq:SMA} was
convolved with the instrumental resolution function 
\cite{CooperNathans,ChesserAxe} and globally fit to scans in the range $0.5\pi 
< \tilde{k} < 0.7\pi $ and $1.3\pi < \tilde{k} < 1.5\pi$, only 
$\mathcal{A}^{\alpha}_{\mathbf{q}}$ was allowed to vary between scans.  
Incoherent background was modeled as a constant plus a Gaussian function 
centered on $\omega = 0$. The broad peak widths and \textbf{Q}-dependence of 
lineshapes visible in \figref{fig:const_Q_scans} (a)-(d) and 
\figref{fig:SQW_maps} (a)-(b) are accounted for entirely by resolution effects 
and two steeply dispersing modes indicating that the excitations are 
long-lived.  Additional intensity appears in $(h,k,2k)$ plane which is best 
accounted for by a damped harmonic oscillator, blue shaded area in 
\figref{fig:const_Q_scans} (e) and (f).  \footnote{The high energy scattering 
    is much broader than the resolution and does not disperse in the 
    \textbf{Q}-range probed. Since this intensity does not appear in scans 
    performed at equivalent \textbf{Q} positions in the $(0,k,l)$ plane it is 
    likely spurious scattering resulting from the coarse out-of-plane 
resolution of the vertically focusing monochrometer.} Dispersion parameters 
determined from the low-energy global fit were: $\Delta_1 =$ 1.90(9) meV, 
$\Delta_2 =$ 3.95(8) meV, $v_1 =$ 16.4(1.4) meV, $v_2=$19.6(1.7) meV, $\delta_1 
=$  0.074(2), $\delta_2=$ 0.053(4). Scans simulated using these parameters are 
shown as solid lines in \figref{fig:const_Q_scans} and the resulting dispersion 
is plotted as solid and broken grey lines in \figref{fig:SQW_maps} (a). 

The incommensurate wavevector of the lower branch $\zeta_1 = 0.574 \pi$ can be 
used to estimate the ratio $J_2/J_1$ in the two limits of strong rung coupling 
$J_4 > J_1$ and uncoupled weakly interacting $J_1$-$J_2$ chains.  For strong 
rung coupling the wavevector minimizing the dispersion can be estimated 
exactly\cite{Lavarelo:11}
\begin{equation}
    \zeta = \arccos\left(\frac{-J_1}{4J_2}\right),
\end{equation}
giving $J_2/J_1  = 1.08$. In the opposite limit of isolated $J_1$-$J_2$ chains,
density matrix renormalization group (DMRG) results \cite{White:96} can be used
to estimate $J_2/J_1$ from $\zeta$ giving $J_2/J_1 \approx 0.88$. Our INS data
situates \BCPO in an intermediate coupling regime; however, the two limits of
isolated chains and strong rung coupling tightly constrain $J_2 \approx J_1$.

Information about the length scales associated with spin correlations can be
extracted from the equal-time structure factor, shown in \figref{fig:SQW_maps} 
(d). It was calculated by numerically integrating constant $\mathbf{q}$ scans 
between 0.5 and 15 meV, all scans were corrected for the isotropic \CuIon 
form-factor prior to integration.\cite{Brown:06} Static correlations in \BCPO 
are well described by a double peaked square root Lorentzian, characteristic of 
spin chains with competing NN and NNN interactions and exponentially decaying 
spin-spin correlations\cite{Nomura:03}
\begin{equation}
    S(k) \propto \frac{1}{\sqrt{\kappa^2 + (k - \zeta')^2 }} +
    \frac{1}{\sqrt{\kappa^2 + (k - \zeta'')^2 }},
    \label{eq:static_struct}
\end{equation}
where $\zeta' = 0.5 + \delta$, $\zeta'' = 1.5 - \delta$ and $\xi = 1/\kappa$ is
the correlation length. Fitting Eq.~\eqref{eq:static_struct} to data in
\figref{fig:SQW_maps} (d) yields a spin-spin correlation length  of  
$\xi/\left(b/2\right)= 8.0(6)$ and $\delta = 0.074(5)$.

The $\sim$ 2 meV bandwidth along the \rvec{c} direction implies that there is a
significant coupling between the ladder units. However, it is not possible to
distinguish either $J_3$ or $J_4$ as the inter/intra ladder coupling from the
shape of the dispersion alone. More information is contained in the modulation 
of INS intensity along the \rvec{c} and \rvec{a} directions. In the absence of 
a microscopic model for the spin dynamics in \BCPO the first moment sum rule 
can be utilized to determine the relative contribution of spin-pair 
correlations across each bond to the ground state 
energy.\cite{HoenbergBrinkman:74,Stone:01} For a Heisenberg Hamiltonian the 
first moment sum rule is written
\begin{align}
    \langle \omega \rangle_{\mathbf{q}} &= \int_{-\infty}^{\infty} \omega
    S^{\alpha\alpha}(\mathbf{q},\omega)d\omega\notag\\
    & =  -\frac{1}{3}\sum_{j,j'}J_{jj'}\langle \mathbf{S}_{j}\mathbf{S}_{j'}
    \rangle \left[ 1 - \cos{(\mathbf{q}\cdot\mathbf{r}_{jj'})}\right],
    \label{eq:sum_rule}
\end{align}
where $\langle S^{\beta}_{j}S^{\beta}_{j'} \rangle $ is the spin-spin
correlation across bond $\mathbf{r}_{jj'}$. Within the single mode 
approximation the dimer-interference term $\left[ 1 -
    \cos{(\mathbf{q}\cdot\mathbf{r}_{jj'})}\right]$ accounts entirely for the
$\mathbf{q}$-dependent intensity modulation of the inelastic scattering
intensity along the rung direction for a system of isolated 
ladders\cite{Masuda:06, Notbohm:07}.  However, the sum rule in 
Eq.~\eqref{eq:sum_rule} is strictly
only valid for a centro-symmetric lattice obeying inversion symmetry, hence must
be applied with caution when analyzing the spectrum from \BCPO where a
significant DM anisotropy on each rung is potentially present. The first-moment 
was measured in \BCPO by numerically integrating constant-$\mathbf{q}$ scans 
along the \rvec{a} and \rvec{c} directions and globally fit to equation 
\eqref{eq:sum_rule}.  All scans were corrected for the isotropic \CuIon 
form-factor prior to integration.\cite{Brown:06} The best fit was obtained with 
$J_{3}\langle \mathbf{S}_0 \mathbf{S}_3 \rangle $ = -12(5)~meV, $J_{4}\langle 
\mathbf{S}_0 \mathbf{S}_4 \rangle $ = -58(2)~meV.  Intensity modulation along 
the \rvec{c} and \rvec{a} directions is well described including only bonds 
$J_3$ and $J_4$ the results of the fit are shown as solid lines in 
\figref{fig:SQW_maps} (e) - (f). The intensity modulation is thus consistent 
with a two-leg ladder formed by antiferromagnetic rung coupling $J_4$ and 
weaker interladder coupling $J_3$.

Our neutron scattering results confirm that the $J_1$-$J_2$-$J_4$ two-leg
ladder with interladder exchange $J_3$ is an appropriate description of
\BCPO.  However, the gap energy differs significantly from the value of 
$\Delta\!\!\sim$\!  2.9~meV reported by previous thermodynamic measurements.  
\cite{Koteswararao:07,Mentre:09,Tsirlin:10} This discrepancy is potentially 
resolved by the presence of strong anisotropic interactions neglected in the 
thermodynamic analysis. Indeed, DM anisotropies are permitted by the 
crystallographic symmetries of \BCPO. These anisotropies can act to split the 
degeneracy of the lowest lying excitations in zero field, so that the
thermodynamic properties are controlled by two gaps. It is notable, if perhaps
coincidental, that the 2.9~meV thermodynamic gap is in agreement with the
average of the two gaps measured by INS, $1/2(\Delta_1 + \Delta_2) = 
2.93(6)$~meV. We have also checked that the low-temperature magnetic specific 
heat can be fit using an effective 1D model including the sum of contributions 
from both the low and high energy modes $C_m = 1/2 \left(g_1\exp(-\Delta_1/T) + 
g_2 \exp(-\Delta_2/T)\right)$ with each gap fixed at the value determined from 
INS. Furthermore, a zero-field splitting of the lowest lying excitation 
resulting from a staggered DM interaction is consistent with the temperature 
dependence data presented next.

\subsection{Temperature Dependence}
\begin{figure}[htb]
    \includegraphics[scale = 0.9]{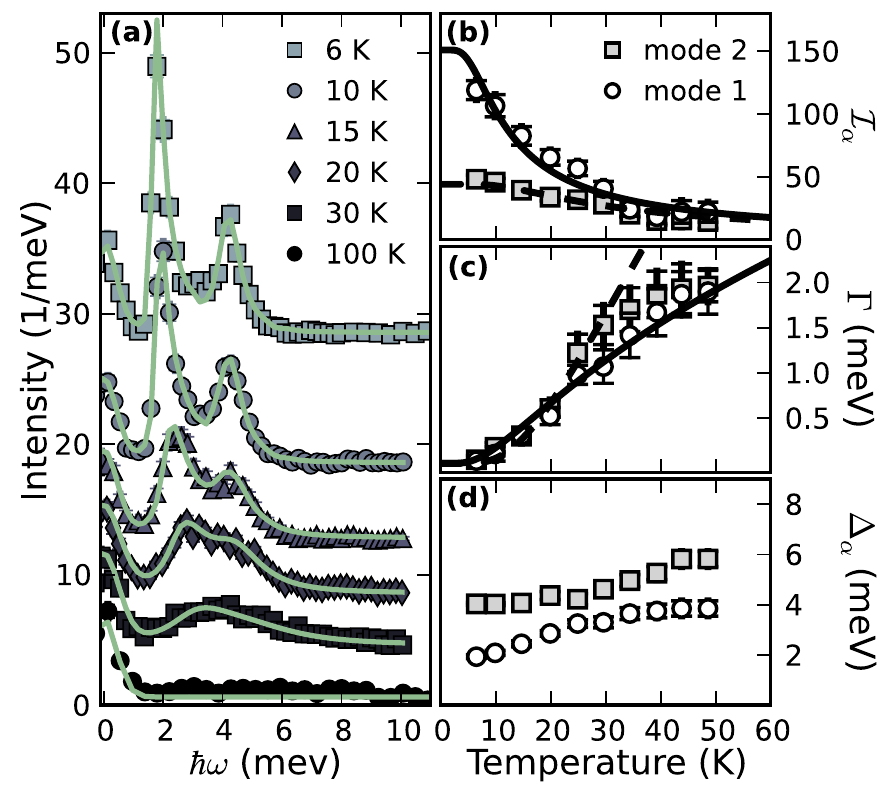}
    \caption{\label{fig:T_dep} Temperature dependence of spin excitations near
    the incommensurate wavevector at (0,1.425,1), data collected on C5. (a) A
    series of representative constant-\textbf{q} scans, data has been offset 
    for clarity, solid lines are fit to resolution convolved cross-section. (b) 
    - (d) Results of fit to resolution convolved Lorentzian scattering 
    cross-section. Solid and dashed lines in (b) are fit to the population 
    thermalization factor from RPA theory. The solid and dashed lines in (c) 
are a fit to exponential activated behaviour as described in the text.}
\end{figure}

The temperature dependence of spin excitations at $\mathbf{q} = (0,1.425,1)$ 
near the incommensurate wavevector is shown in \figref{fig:T_dep}(a). As the 
temperature is increased the inelastic features broaden and shift to higher 
energies. Above 25~K the two excitations can no longer be distinctly identified 
and above 55~K the inelastic signal cannot be distinguished from background.
Each spectrum was fit to the resolution convolved single mode cross
section [Eq.\! \eqref{eq:SMA}] with a normalized Lorentzian function in place of
the delta function to account for intrinsic broadening of each mode.  Lacking
detailed measurements of the dispersion for $T > 6$~K the spin
wave velocity was assumed to be independent of temperature for all fits.
Constant-\textbf{q} scans were performed at a wavevector where the orientation
of the resolution ellipsoid with respect to the slope of the dispersion
minimizes any resolution effects. We have also checked that allowing for a
temperature dependent spin wave velocity does not affect the results presented
here. Figures \ref{fig:T_dep} (b) - \ref{fig:T_dep}(c) show the temperature
dependence of the gap energy, half-width half-maximum and integrated intensity
of each mode.

In accordance with the thermalization of the ground and excited states, the
integrated intensity decreases with increasing temperature. The random phase
approximation (RPA) for weakly interacting dimers predicts the scattering
intensity to scale with the ground-state excited-state population difference
$\Delta n(\Delta_{\alpha}/T) = \left(1 - \exp \left( -\Delta_{\alpha} / {T}
\right) \right) /\left( 1 + \eta \exp\left(-\Delta_{\alpha}/T\right)\right )$,
where $\eta$ is the degeneracy of the level.
\cite{Luenberger:84,Matsuda:99,Xu:00} The RPA form is a good description of the
data after fitting only an overall scale factor [solid lines in
\figref{fig:T_dep} (b)]. The best fit is obtained assuming the two modes are
composed of a lower energy doubly degenerate excitation ($\Delta_1$) and a
higher energy singly degenerate excitation ($\Delta_2$).  Although we do not
have sufficiently detailed data to completely rule out the possibility of two
triply degenerate modes. The scaling of the intensities provides good
qualitative evidence for the splitting of a single triply degenerate excitation
band into a doublet and singly degenerate excitation.

For systems with a gapped excitation spectrum increasing temperature results in
an increased density of thermally activated states and concomitantly a decrease
in the mean free path between scattering of the excited states, or equivalently
a decreased excitation lifetime.\cite{Damle:98,Zheludev:08,Nafradi:11} Thus, we 
expect a thermally induced broadening of the excitation spectrum, as visible in
\figref{fig:T_dep} (c). Below 30 K the damping is consistent with an 
exponential activated behaviour $\Gamma_{\alpha} = g_{\alpha} 
\sqrt{\Delta_{\alpha}/T} \exp{(\Delta_{\alpha}/T)}$, valid for gapped
one-dimensional systems.\cite{Damle:98} The fitted scale factors were $g_1 =
1.05(1)$ and $g_2 = 2.22(16)$ for the low and high energy modes respectively
and the fits are shown as solid and dashed lines in \figref{fig:T_dep}(c). A
slight increase in excitation energy with increasing temperature was also
observed [\figref{fig:T_dep} (d)]. This is understood in the context of the
thermally induced ''blue-shift'' that has been reported in other
quasi-one-dimensional systems.
\cite{Zheludev:96,Kenzelmann:01,Xu:07,Zheludev:08,Nafradi:11}.

\section{Discussion}
The excitation spectrum in \BCPO is unique in that competing interactions along
the spin chains drive the dynamic correlations to an incommensurate 
wave-vector. Incommensurate dynamic correlations have been observed in the 
four-leg antiferromagnetic spin tube $\mathrm{Sul-Cu}_2\mathrm{Cl}_4$ 
\cite{Garlea:08}; however, the degree of incommensurability was much smaller 
than that observed in \BCPO\!\!. Furthermore, the excitations in 
$\mathrm{Sul-Cu}_2\mathrm{Cl}_4$ are comprised of a single, triply degenerate 
mode, while in \BCPO the degeneracy is split and the minimum of each mode 
occurs at a different wave vector.

In a canonical two-leg ladder the low energy magnetic excitation spectrum is
composed of a well defined branch of dispersing triplets. Application of an
external magnetic field splits the degeneracy and three branches of excitations
may be observed.\cite{Hong:10} The degeneracy may also be split in the
presence of anisotropic interactions, such as the DM anisotropy, in the
microscopic Hamiltonian.\cite{Miyahara:07,Wang:02} We believe that the
observed mode splitting in \BCPO may be accounted for by significant staggered 
DM interactions. In \BCPO local inversion symmetry is broken allowing a DM
interaction, with the particular crystallographic symmetries admitting a DM
anisotropy with a staggered DM vector on each rung as well as DM interactions
on the chain bonds.\cite{Tsirlin:10}  Although there are many possible DM
vectors, our INS measurements can only resolve two-bands of excitations in
zero-field indicating that the magnetic Hamiltonian retains a local symmetry,
so that we expect a single, dominant, anisotropy term. Indeed DM interactions
account for many features of the thermodynamic measurements including the weak
linear field dependence of the low field magnetization below H$_\mathrm{c}$
\cite{Tsirlin:10,Kohama:12}, highly anisotropic critical fields
\cite{Kohama:12}, and absence of square root singularity in magnetization just
above H$_\mathrm{c}$. \cite{Tsirlin:10,Miyahara:07}  Field theoretical modeling
has shown that staggered anisotropic interactions in a spin-1/2 two-leg ladder
can act to split the lowest lying triplet into a doublet with energy $\Delta_d
< \Delta_t$ and higher energy mode with $\Delta_3 > \Delta_t$.\cite{Wang:02}
Our temperature dependent measurements are consistent with this scenario,
although further detailed magnetic field dependent measurements are required to
definitively establish the degeneracy of each mode.

In summary, we have measured the spin excitation spectrum in \BCPO at 6~K. The 
results confirm that \BCPO is in a quantum disordered phase and that the 
$J_1$-$J_2$-$J_4$ two-leg ladder model with interladder exchange $J_3$ 
appropriately accounts for the dominant exchange pathways. The measured 
incommensurate wavevector  $\zeta = 0.574\pi$ allows for an estimate of the 
relative strength of the competing magnetic interactions $J_2/J_1 \sim 1$.  
However, the lowest excitation gap is 1.90(9), is significantly lower than the 
value of 2.9 predicted by thermodynamic measurements. The discrepancy may be 
explained by the presence of strong anisotropic magnetic interactions which 
split the excitation spectrum into two coherent branches. We hope that our data 
will stimulate further theoretical work investigating the effect of 
Dzyloshinsky-Moria interactions and frustration on the spin-liquid ground
state of a even-leg ladder.

\begin{acknowledgments}
We would like to thank Yong-Baek Kim, Arun Paramekanti, and Leon Balents for
useful discussions. Y.J. Kim acknowledges the hospitality and the Aspen Center
for Physics supported in part by the National Science Foundation under grant No.
PHYS-1066293. Work at the University of Toronto was supported by NSERC of
Canada.  Work at Chalk River Labs was supported by NSERC of Canada, NRC of
Canada.  Work at HFIR was sponsored by the Scientific User Facilities Division,
Office of Basic Energy Sciences, U.S.  Department of Energy.  K.W.  Plumb
acknowledges the support of the Ontario Graduate Scholarship.
\end{acknowledgments}

\end{document}